\documentclass[aps,preprintnumbers,twocolumn,amsmath,superscriptaddress,amssymb,floatfix,pra]{revtex4-1}
\pdfoutput=1
\usepackage[utf8]{inputenc}
\usepackage{amsmath}
\usepackage{amsfonts}
\usepackage{braket}
\usepackage{tensor}
\usepackage{graphicx}
\usepackage[colorlinks=true,linkcolor=blue, citecolor=blue, urlcolor=blue, bookmarks]{hyperref}

\def\be{\begin{equation}}
\def\ee{\end{equation}}
\def\bea{\begin{eqnarray}}
\def\eea{\end{eqnarray}}

\def\XXint#1#2#3{{\setbox0=\hbox{$#1{#2#3}{\int}$}
         \vcenter{\hbox{$#2#3$}}\kern-.5\wd0}}

\begin{document}

\title{Correlations and diagonal entropy after quantum quenches in XXZ chains}
\author{Lorenzo Piroli}
\author{Eric Vernier}
\author{Pasquale Calabrese}
\affiliation{SISSA and INFN, via Bonomea 265, 34136 Trieste, Italy.}
\author{Marcos Rigol}
\affiliation{Department of Physics, The Pennsylvania State University, University Park, PA 16802, USA}

\begin{abstract}
We study quantum quenches in the XXZ spin-$1/2$ Heisenberg chain from families of ferromagnetic and antiferromagnetic initial states. Using Bethe ansatz techniques, we compute short-range correlators in the complete generalized Gibbs ensemble (GGE), which takes into account all local and quasi-local conservation laws. We compare our results to exact diagonalization and numerical linked cluster expansion calculations for the diagonal ensemble finding excellent agreement and thus providing a very accurate test for the validity of the complete GGE. Furthermore, we use exact diagonalization to compute the diagonal entropy in the postquench steady state. We show that the Yang-Yang entropy for the complete GGE is consistent with twice the value of the diagonal entropy in the largest chains or the extrapolated result in the thermodynamic limit. Finally, the complete GGE is quantitatively contrasted with the GGE built using only the local conserved charges (local GGE). The predictions of the two ensembles are found to differ significantly in the case of ferromagnetic initial states. Such initial states are better suited than others considered in the literature to experimentally test the validity of the complete GGE and contrast it to the failure of the local GGE.
\end{abstract}

\maketitle

\section{Introduction}\label{intro}

Understanding the long-time behavior of an isolated many-body quantum system after it is brought out of equilibrium represents a fundamental physical problem. Arguably, the simplest paradigm is that of a quantum quench \cite{cc-06}, where a well-defined initial state is let evolved unitarily under a time independent Hamiltonian.

In the past decade, tremendous theoretical efforts have been devoted to the study of quantum quenches, an important motivation being recent experimental progress in ultracold atomic physics \cite{bdz-08,ccgo-11,pssv-11}. Many beautiful experiments have now shown that nearly isolated quantum systems can be taken far from equilibrium using quantum quenches and their dynamics studied in great detail \cite{kww-06, cetal-12, gklk-12, shr-12, mmk-13, fse-13, lgkr-13, mmkl-14, adrl-14, glms-14, ktlr-16}. Such experiments have made it clear the need for a theoretical understanding of quantum dynamics in isolated systems, which may have been considered to be a purely academic problem before. 

Among others, one question has emerged as especially important: is it possible to predict the properties of a post-quench steady state based on simple physical principles? This question is particularly relevant when compared to the prohibitive complexity of computing the full post-quench time evolution of a many-body quantum system, and has inspired an intense theoretical research activity. As a result, it has been concluded that a fundamental difference exists between generic and integrable systems. In the former, it was proposed that the post-quench steady state can be described by a thermal Gibbs ensemble, where the effective temperature is fixed by the initial state \cite{deutsch-91, rdo-08, dkpr-16, efg-15, goei-16}. On the other hand, in integrable models a key role is played by the presence of an extensive number of conservation laws and such systems retain much more information on the initial state \cite{rf-11,heri-12}. Accordingly, in order to characterize the post-quench steady state a generalized Gibbs ensemble (GGE) was proposed, which is constructed by taking into account conservation laws emerging from integrability \cite{rdyo-07,rmo-06,ccr-11}. 

Much subsequent work has addressed the subtle question of which conserved operators (or charges) have to be taken into account in the GGE. In integrable models mappable to noninteracting ones, either the occupation of the single-particle eigenstates of the noninteracting problem or all {\it local} conserved charges have been shown to produce correct physical predictions following quantum quenches \cite{rdyo-07, rmo-06, ccr-11, cef-11, fe2-13, cazalilla-06, cdeo-08, ic-09, fm-10, ce-10, eef-12, sfm-12, mc-12, ck-12, gr-12, gurarie-13, csc-13, kcc-14, rs-14, wrdk-14, sc-14, mckc-14, bkc-14, rs-15, sm-16, gkf-16}. However, recent investigations in models that cannot be mapped onto noninteracting ones (referred to as interacting integrable models in what follows) have unraveled the need for considering more generally all {\it quasi-local} conservation laws \cite{fe-13, pozsgay-13, fcec-14, wdbf-14, pmwk-14, pozsgay2-14, ga-14, idwc-15, imp-15, iqdb-16,iqc-16}. 

The existence of quasi-local conserved charges in interacting integrable lattice models was discovered in the context of thermal spin transport \cite{prosen-11}, where a number of works have been devoted to analyze their physical consequences \cite{ip-12, prosen-14, ppsa-14, pv-16, zmp-16}. Note that additional non-local conservation laws and their importance on relaxation processes have also been recently discussed in XY spin chains \cite{fagotti-14, bf-15, fc-15, fagotti-16, fagotti2-16} and integrable quantum field theories \cite{emp-15, bs-16, dlsb-15, vecu-16, cardy-16}. The ensemble obtained by considering only local conserved operators is now called the local (or ultra-local) GGE, while the correct construction which includes all quasi-local charges is usually referred to as the complete GGE. For a pedagogical introduction to these topics see the recent reviews \cite{cem-16, ef-16, impz-16, vr-16, dm-16, cc-16}.

An important role in the more recent developments has been played by the introduction of the so-called quench action approach \cite{ce-13,caux-16}. The latter is an analytical method that allows one to compute physical quantities on the post-quench steady state based on Bethe ansatz techniques. At the moment, its applicability is limited to those initial states for which the overlaps with the eigenstates of the Hamiltonian after the quench are known analytically \cite{fcc-09, mpc-10, dwbc-14, pozsgay-14, bdwc-14,pc-14, msca-16,ppv-16}. Nevertheless, it has already been successfully employed in the study of several quantum quenches \cite{dwbc-14,  bse-14, dc-14, dmv-15, pce-16, alca-16, wdbf-14, pmwk-14}, and was an essential theoretical tool for establishing the failure of the local GGE in interacting systems \cite{wdbf-14,pmwk-14}.

Most of the progress related to interacting integrable models has occurred remarkably rapidly. It is then imperative to substantiate the body of evidence in favor of the complete GGE conjecture, which at the moment has been tested to high precision for a relatively small number of initial states \cite{wdbf-14, pmwk-14, pvc-16}. Furthermore, in the majority of the cases, the local GGE has been found to provide extremely good predictions for local observables. An important question related to experiments is whether one can always expect the local GGE to give predictions that are sufficiently accurate for all practical purposes. 

The systematic investigation of these issues provides the first motivation for our work. In particular, we focus on the prototypical XXZ spin-$1/2$ Heisenberg chain and consider families of initial states given by antiferromagnets and tilted ferromagnets for several values of the tilting angle. Complete GGE results for some tilting angles and local correlators involving those families of initial states have been presented in Ref.~\cite{pvc-16}. Here, we systematically compare the complete GGE against calculations for the diagonal ensemble using exact diagonalization (ED) and numerical-linked cluster expansions (NLCEs) \cite{rigol-14, rigol-14a, rigol-16}. We find excellent agreement in most cases, and, in those in which the results do not agree, we find that the exact diagonalization and NLCE calculations approach the complete GGE results with increasing the chain and the cluster sizes, respectively. Thus, our study provides the most accurate benchmark to date of the complete GGE. Furthermore, we show that for tilted ferromagnets the predictions of the local and complete GGEs are significantly different, only the latter being in agreement with ED and NLCE calculations. This observation likely shifts the failure of the local GGE from a purely academic result to something that can be effectively tested experimentally.

We also characterize the post-quench steady states beyond short-distance correlators. In the Bethe ansatz language, any statistical ensemble (and hence any GGE) can be characterized in terms of quasi-particle rapidity distribution functions, which generalize the concept of particle momentum distribution from noninteracting systems. In fact, one of the simplest quantities in this context is provided by the corresponding entropy, the so-called Yang-Yang entropy \cite{yy-69}. While it is an established result that for thermal states the latter coincides with the thermal entropy \cite{takahashi-99}, its meaning for non-equilibrium steady states has yet to be clarified. We note that the Yang-Yang entropy has recently been used as one of the key ingredients in the computation of entanglement dynamics in Heisenberg spin chains \cite{vc-16}, and has thus already proven to be of great interest in the study of quantum quenches. A detailed investigation of this quantity provides the second motivation of our work.

In particular, we study the relation between the Yang-Yang entropy and the so-called diagonal entropy, whose thermodynamic meaning has been discussed in Refs.~\cite{rk-06, polkovnikov-11,spr-11} (see Ref.~\cite{dkpr-16} for a recent review). For all the initial states considered, we show that the Yang-Yang entropy for the complete GGE is consistent with twice the value of the diagonal entropy in the largest chains or the extrapolated result in the thermodynamic limit. This is similar to the results obtained for several quantum quenches in translationally invariant systems that are either noninteracting or that can be mapped onto noninteracting ones \cite{gurarie-13, ckc_2-14, fc-08} (see Ref.~\cite{vr-16} for a recent review). Our findings are also in agreement with the analysis of Ref.~\cite{vc-16} and corroborate the picture of pair quasi-particle production after a quantum quench even in fully interacting integrable models such as the XXZ spin-$1/2$ Heisenberg chain.

The presentation is organized as follows. In section~\ref{sec:setup}, we introduce the XXZ spin-$1/2$ Heisenberg Hamiltonian and the quantum quenches considered in this work, while in section~\ref{sec:bethe_ansatzs} we briefly introduce the Bethe ansatz language used to describe the complete GGE. In section~\ref{sec:local_correlations}, we compare the Bethe ansatz predictions with the results from ED and NLCEs for the diagonal ensemble. Section~\ref{sec:diagonal_entropies} is devoted to the analysis of diagonal and Yang-Yang entropies. Our conclusions are reported in section~\ref{sec:conclusions}. Technical details from our calculations are reported in the appendices.

\section{Hamiltonian and quantum quenches}\label{sec:setup}
The XXZ spin-$1/2$ Heisenberg Hamiltonian can be written as
\bea
H&=& \frac{1}{4}\sum_{j=1}^{L}\left[\sigma^{x}_j\sigma^{x}_{j+1}+\sigma^{y}_j\sigma^{y}_{j+1}+\Delta\left( \sigma^{z}_j\sigma^{z}_{j+1}-1\right)\right] \,,
\label{eq:hamiltonian}
\eea
where $\sigma_j^{\alpha}$, $\alpha=x,y,z$, are the Pauli matrices, and $\Delta$ is the anisotropy parameter. Later, we will use $\sigma^{\pm}=(\sigma^x\pm i\sigma^y )/2$. We restrict our study to the regime $\Delta\geq 1$. 

We consider quantum quenches from two different initial states, namely, the N\'eel state
\be\label{eq:neel}
|N\rangle=|\uparrow\downarrow\rangle\otimes\ldots\otimes |\uparrow\downarrow\rangle\,=|\uparrow\downarrow\rangle^{\otimes L/2},
\ee
and the tilted ferromagnet
\be
|\Theta;\nearrow\rangle=\left[\cos\left(\frac{\Theta}{2}\right)|\uparrow\rangle+i\sin\left(\frac{\Theta}{2}\right)|\downarrow\rangle\right]^{\otimes L}\,.
\label{eq:ferromagnet}
\ee
These states are ground states of simple Hamiltonians. The N\'eel state is the ground state of Hamiltonian~\eqref{eq:hamiltonian} for $\Delta\rightarrow\infty$ (the antiferromagnetic Ising model), or of $H=\sum_{j=1}^{L}(-1)^j\sigma^{z}_j$. While the tilted-ferromagnet is the ground-state of Hamiltonian~\eqref{eq:hamiltonian} in the presence of a very strong magnetic field pointing in the tilting direction.
Quantum quenches from the N\'eel state were previously considered in Refs.~\cite{wdbf-14, pmwk-14}, were the quench action approach was employed to obtain an explicit characterization of the post-quench steady state. In Ref.~\cite{idwc-15}, it was shown that the latter coincides with the complete GGE constructed using all local and quasi-local charges. For the tilted ferromagnet, in contrast, the overlaps needed to implement the quench action are not known and the construction of the complete GGE provides for the moment the only available analytical approach to obtain predictions on the post-quench steady state. 

In the next section, we review the Bethe ansatz description of the complete GGE in terms of rapidity distribution functions.

On the other hand, in our numerical calculations using ED and NLCEs (see appendix~\ref{sec:diag_nlces}), correlators after the quench are computed in the so-called diagonal ensemble \cite{rdo-08}, for which the density matrix reads
\be\label{eq:diagensem}
\rho_{\text{DE}}\equiv\lim_{t'\rightarrow\infty}\frac{1}{t'}\int_0^{t'} 
dt\,\rho(t)=\sum_\alpha W_\alpha\, |\alpha\rangle\langle\alpha|,
\ee
where $\rho(t)$ is the time-evolving density matrix after the quench, $|\alpha\rangle$ are the eigenstates of the final Hamiltonian, and $W_\alpha$ are the weights of the initial state in the eigenstates of the final Hamiltonian. In Refs.~\cite{wdbf-14, rigol-14a}, NLCEs were used to compute correlators in the diagonal ensemble after quenches from the N\'eel state. The results obtained were in excellent agreement with those from the quench action approach.

\section{Complete GGE from Bethe ansatz}\label{sec:bethe_ansatzs}
The Hamiltonian \eqref{eq:hamiltonian} is integrable and can be diagonalized by means of the Bethe ansatz \cite{takahashi-99, kbi-93}. A general eigenstate has a well defined number $M$ of down spins and can be written as
\bea
\left|\{\lambda_j\}_{j=1}^M\right\rangle&=&\sum_{x_1<\ldots<x_M}\sum_{Q\in S_M}A_{Q}\left(\{\lambda_j\}_{j=1}^M\right)\nonumber\\
&\times & \prod_{j=1}^Me^{-ix_jp(\lambda_{Q_j})}\sigma^{-}_{x_j}|\uparrow\ldots\uparrow\rangle\,,
\label{eq:wave_function}
\eea
where
\bea
p(\lambda)&=&-i\ln\left[\frac{\sin(\lambda+i\eta/2)}{\sin(\lambda-i\eta/2)}\right]\,,\\
A_{Q}\left(\{\lambda_j\}\right)&=&\prod_{k<j}^M\frac{\sin(\lambda_{Q_j}-\lambda_{Q_k}-i\eta)}{\sin(\lambda_{Q_j}-\lambda_{Q_k})}\,,
\eea
where $\eta={\rm arccosh}(\Delta)$. The second sum in Eq.~\eqref{eq:wave_function} is over all permutations of $M$ elements.

The complex parameters $\{\lambda_j\}_{j=1}^M$, usually called rapidities, are obtained as the solution of so-called Bethe equations
\bea
\left(\frac{\sin(\lambda_j+i\eta/2)}{\sin(\lambda_j-i\eta/2)}\right)^{L}=-\prod_{k=1}^{M}\frac{\sin(\lambda_j-\lambda_k+i\eta)}{\sin(\lambda_j-\lambda_k-i\eta)}\,.
\label{eq:bethe_eq}
\eea
The energy of an eigenstate corresponding to the set $\{\lambda\}_{j=1}^M$ is then given by
\bea
e\left[\{\lambda_j\}_{j=1}^M\right]&=&-\sum_{j=1}^M\frac{\sinh^2\eta}{\cosh(\eta)-\cos(2\lambda_j)}\,.
\eea
The solutions of Eqs.~\eqref{eq:bethe_eq} arrange themselves into patterns in the complex plane called strings. A solution $\{\lambda_j\}_{j=1}^M$ consists of $M_n$ strings of length $n$ in which the rapidities are parametrized as
\be
\lambda_{\alpha}^{n,a}=\lambda_{\alpha}^{n}+i\frac{\eta}{2}(n+1-2a)+i\delta_{\alpha}^{n,a}\,,
\ee
with $a=1,\ldots n$. Here, the real numbers $\lambda_{\alpha}^n$ are called string centers and satisfy $\lambda_{\alpha}^n\in [-\pi/2,\pi/2]$, while $\delta_{\alpha}^{n,a}$ are exponentially small deviations that are ignored in the thermodynamic limit, within the so-called string hypothesis \cite{takahashi-99}.

From the wave-function \eqref{eq:wave_function}, magnonic excitations (down spins) can be interpreted as quasi-particles, while $n$-strings can be interpreted as bound states of $n$ quasi-particles. This picture provides the basis for the thermodynamic description of the model.

In the thermodynamic limit, the macro-states of the system are described by the quasi-particle and bound-state rapidity distribution functions. In particular, $n$-string centers become dense in the interval $[-\pi/2,\pi/2]$ according to a set of distribution functions $\rho_n(\lambda)$, which completely characterize a macro-state. Together with these, one has distribution functions $\rho_n^h(\lambda)$ for the so-called $n$-string holes, which generalize the concept of holes from a noninteracting Fermi gas. In the interacting model considered here, the functions $\rho_n^h(\lambda)$ are non-trivially related to $\rho_n(\lambda)$ through the thermodynamic version of the Bethe equations \eqref{eq:bethe_eq}
\be
\rho_{n}(\lambda)+\rho^h_{n}(\lambda)=a_n(\lambda)-\sum_{m=1}^{\infty}\left(a_{nm}\ast\rho_m\right)(\lambda)\,,
\label{eq:thermo_bethe}
\ee
where
\bea
a_{nm}(\lambda)&=&(1-\delta_{nm})a_{|n-m|}(\lambda)+2a_{|n-m|}(\lambda)\nonumber\\
 &+&\ldots +2a_{n+m-2}(\lambda)+a_{n+m}(\lambda)\,,
\eea
and
\be
a_n(\lambda)=\frac{1}{\pi} \frac{\sinh\left( n\eta\right)}{\cosh (n \eta) - \cos( 2 \lambda)}\,.
\label{def:a_function}
\ee
In Eq.~\eqref{eq:thermo_bethe}, the convolution between two functions is defined as
\be
\left(f\ast g\right)(\lambda)=\int_{-\pi/2}^{\pi/2}{\rm d}\mu f(\lambda-\mu)g(\mu)\,.
\label{eq:convolution}
\ee

As already mentioned, this thermodynamic formalism can be employed to describe the statistical ensembles provided by the local and complete GGE. In particular, a recent success has been the determination of the rapidity distribution functions corresponding to the complete GGE for the N\'eel and tilted ferromagnetic states considered here \cite{wdbf-14,pmwk-14,iqdb-16,pvc-16}. These are briefly reviewed in appendix \ref{sec:app_distributions}, where we also discuss how the rapidity distribution functions for the local GGE are determined.

In principle, the rapidity distribution functions of a macro-state allow one to compute all local correlators. More precisely, such correlators can be obtained after the numerical solution of sets of non-linear integral equations, which depend on the rapidity distribution functions. Further details are reported in appendix~\ref{sec:app_correlators}, while we refer the reader to the literature for a more complete treatment \cite{wdbf-14,pmwk-14,mp-14,pozsgay-16}.

\section{Short-range correlators}\label{sec:local_correlations}

In this section, we compare the Bethe ansatz predictions with ED and NLCE calculations in the diagonal ensemble. In what follows, we first revisit the N\'eel state, which was previously studied in Refs.~\cite{wdbf-14,pmwk-14,rigol-14a}. Our analysis goes beyond those works in that we consider transverse correlators ($\sigma^{+}_i\sigma^{-}_{i+k}$). Second, we study the tilted ferromagnet for a large number of tilting angles.

\subsection{N\'eel state}

\begin{figure}[!t]
\includegraphics[width=0.48\textwidth]{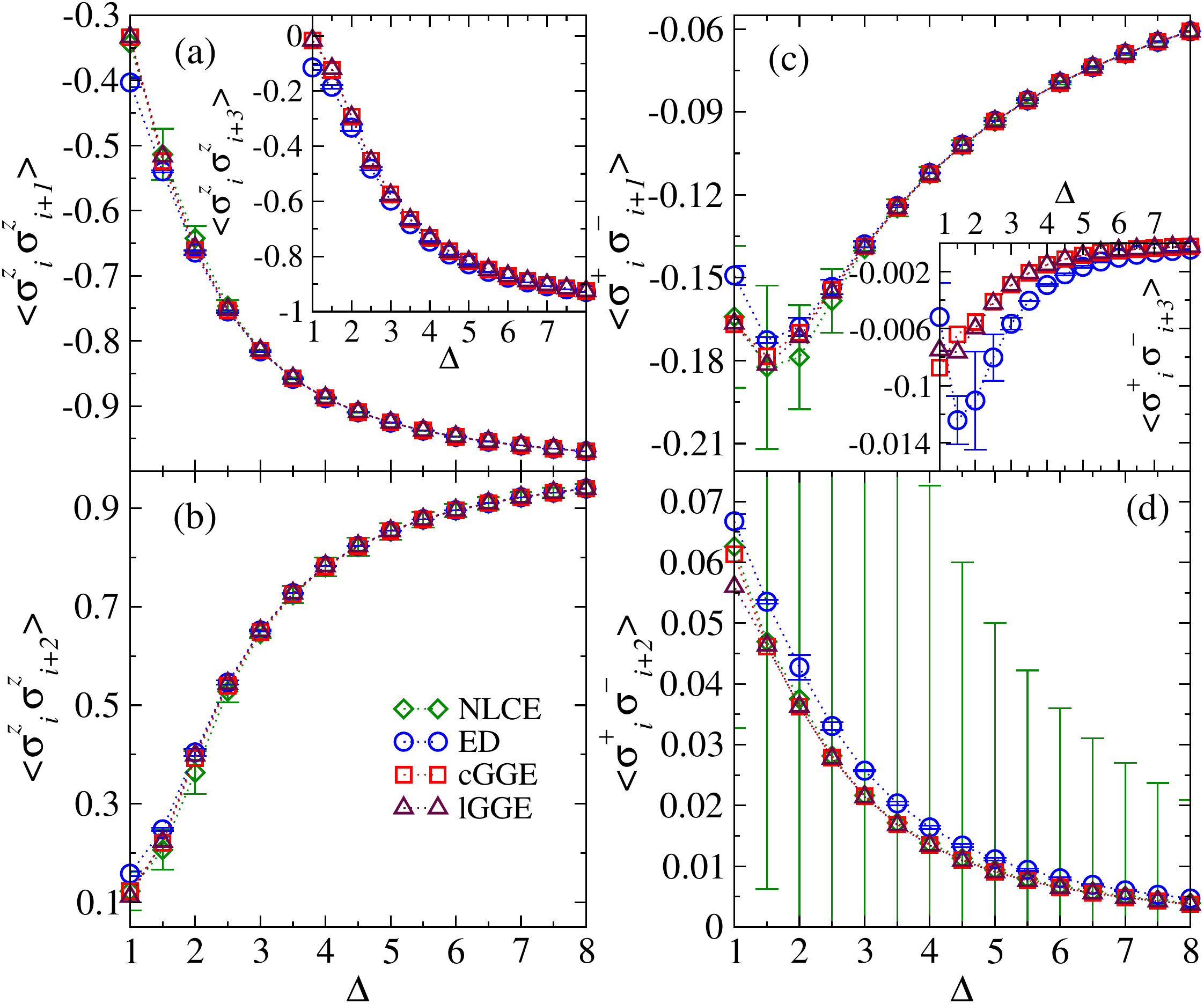}
\caption{Short-range correlators for the quench from the N\'eel state as functions of the anisotropy parameter $\Delta$. The complete GGE (cGGE) and local GGE (lGGE) predictions are compared to the results from ED and NLCE. The ED and NLCE results reported are the average over the two largest chains (with 22 and 24 sites) and the two highest orders of the expansion (18 and 19 orders), respectively. The values at the extremes of the errorbars depict the results that entered the averages. For $\langle\sigma^{z}_i\sigma^{z}_{i+3}\rangle$, inset in (a), and $\langle\sigma^{+}_i\sigma^{-}_{i+3}\rangle$, inset in (c), we only report results for cGGE, lGGE, and ED.}
\label{fig:1}
\end{figure}

Here, we compare the Bethe ansatz predictions for the complete GGE for longitudinal and transverse correlators to the results obtained for the diagonal ensemble from ED in chains with up to 24 lattice sites (with periodic boundary conditions) and NLCEs in clusters with up to 19 sites. Our findings are reported in Fig.~\ref{fig:1}. 

In Fig.~\ref{fig:1}, the results reported from the ED calculations are the average between those obtained in chains with $L=22$ and $L=24$ sites. The actual ED results for the chains with $L=22$ and $L=24$ sites are shown as the extremes of the errorbars. This allows one to see how rapidly the ED results are changing with increasing the chain size. Results are only reported for chains with an even number of sites as those are the only ones that accommodate the N\'eel state (see appendix~\ref{sec:diag_nlces}). From the NLCE calculations, the results reported are the average between those obtained in the expansions with up to 18 and 19 site clusters. The actual NLCE results for up to 18 and 19 site clusters are shown as the extremes of the errorbars. This allows one to gauge convergence in the NLCE calculations.

\begin{figure}[!b]
\includegraphics[width=0.48\textwidth]{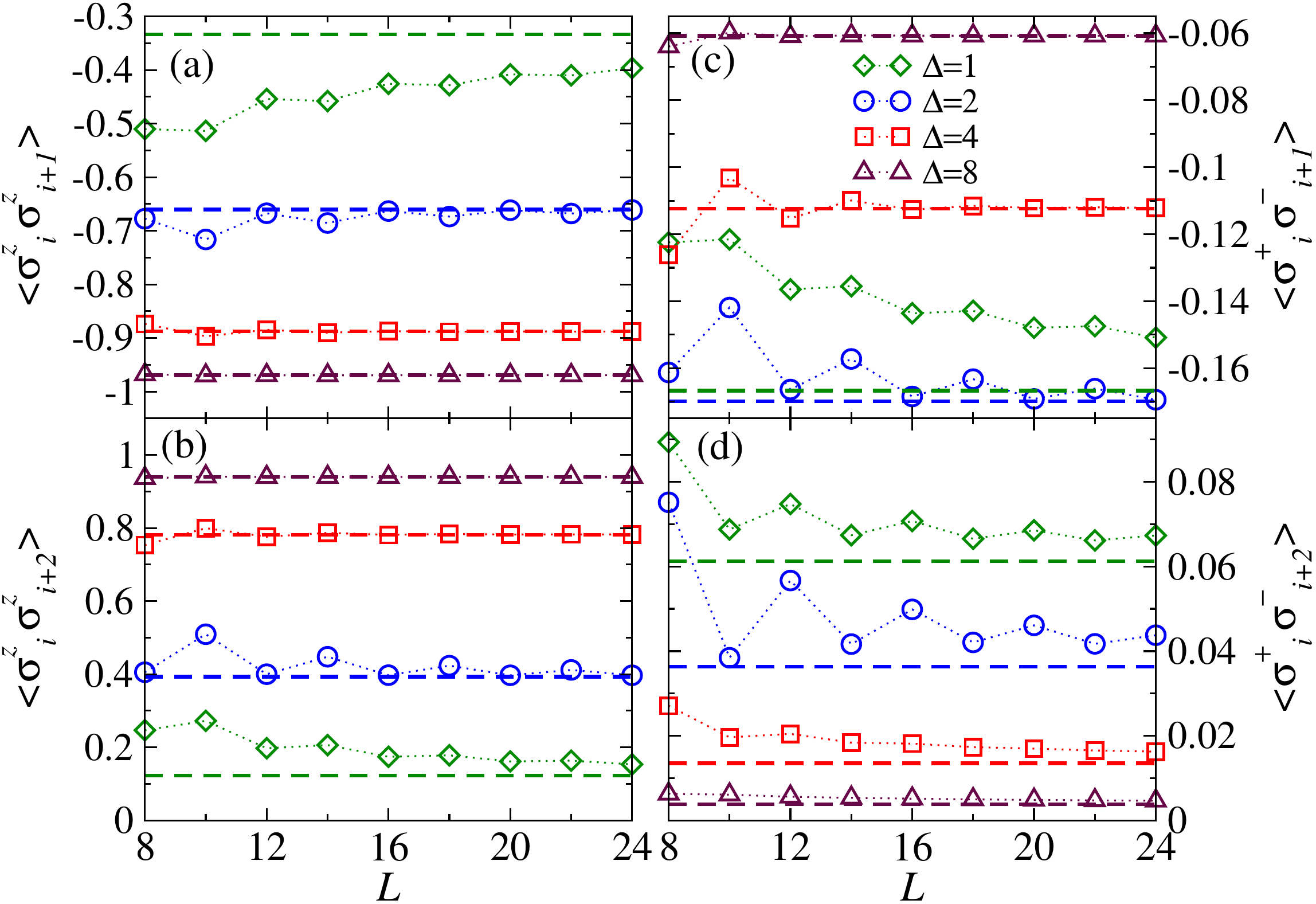}
\caption{Short-range correlators for the quench from the N\'eel state for anisotropy parameters $\Delta=1,$ 2, 4, and 8. The ED results (symbols) are shown as functions of the chain size $L$ in systems with periodic boundary conditions. The complete GGE results are shown as horizontal dashed lines. For all correlators and values of $\Delta$, one can see that the ED results approach the Bethe ansatz predictions as $L$ increases.}
\label{fig:2}
\end{figure}

Figure~\ref{fig:1} shows that, for all the short-range correlators and values of the anisotropy parameter $\Delta$ that we have considered, there is an excellent agreement between the complete GGE and the ED [except for $\langle\sigma^{+}_i\sigma^{-}_{i+2}\rangle$ in (d) and $\langle\sigma^{+}_i\sigma^{-}_{i+3}\rangle$ in the inset in (c), because of finite size effects] and NLCE results. We note that finite-size errors in ED and convergence errors in NLCE increase with decreasing $\Delta$, and with increasing the support of the correlators. However, in all cases, the complete GGE results are within the results for the last two orders of the NLCE expansion, and, in most cases, coincide with their average. The NLCE results fluctuate about the Bethe ansatz prediction and, as shown in Ref.~\cite{rigol-14a}, the magnitude of the fluctuations decrease with increasing the cluster sizes. 

In Fig.~\ref{fig:2}, we show ED results for four values of $\Delta$ as a function of the chain size, and compare them to the complete GGE predictions (horizontal dashed lines). The comparison makes apparent that the ED results approach the complete GGE ones with increasing the chain size. Also, as mentioned in the context of Fig.~\ref{fig:1}, Fig.~\ref{fig:2} shows that finite-size errors increase when decreasing the anisotropy parameter, they are most severe at $\Delta=1$, and with increasing the support of the correlators. 

\begin{figure}[!t]
\includegraphics[width=0.47\textwidth]{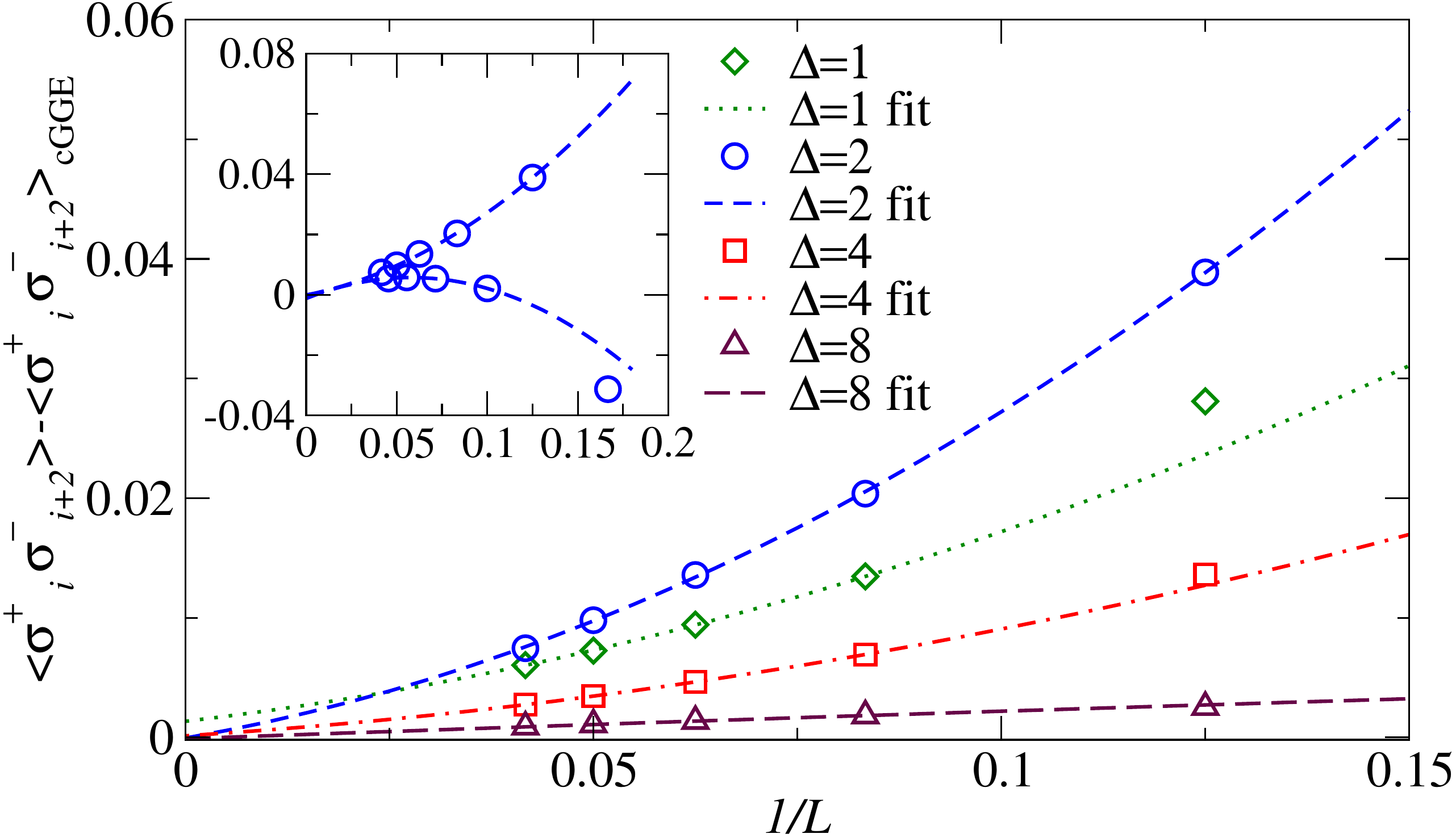}
\caption{Finite-size scaling analysis for $\langle \sigma_j^{+}\sigma_{j+2}^{-}\rangle$ [see also Fig.~\ref{fig:2}(d)]. Main panel: Differences between ED results in chains with $L=4n$ ($n$ integer) and the complete GGE predictions vs $1/L$ (symbols), and their fits to $a+b/L+c/L^2$ (lines). Inset: Comparison between the difference in the main panel for $\Delta=2$ in chains with $L=4n$ and $L=4n+2$. The fits yield identical results as $L\to \infty$.}
\label{fig:2b}
\end{figure}

A detailed analysis of finite-size effects for $\langle\sigma^{+}_j\sigma^{-}_{j+2}\rangle$ [corresponding to the results reported in Fig.~\ref{fig:2}(d)] is undertaken in Fig.~\ref{fig:2b}. Due to even-odd effects, it is necessary to deal with chains of length $L=4n$ and $L=4n+2$, with $n$ integer, separately. The main panel in Fig.~\ref{fig:2b} shows fits ($a+b/L+c/L^2$) to the difference between the ED results in chains of length $L=4n$ and the complete GGE predictions. Our results are consistent with a vanishing value of $a$ (the largest error is obtained for $\Delta=1$). In the inset, we compare fits in chains with $L=4n$ and $L=4n+2$ for $\Delta=2$. They can be seen to predict the same results as $L\to \infty$. 

In Fig.~\ref{fig:1}, we also report the results obtained for the local GGE. They are almost indistinguishable from those obtained for the complete GGE. This was first observed in Ref.~\cite{wdbf-14}, where a detailed analysis for longitudinal correlators was provided. In particular, it was shown that the large-$\Delta$ expansions for the short-range correlators of the local GGE and the complete GGE coincide up to the fourth order and do not differ significantly in general \cite{wdbf-14}. On the other hand, our results for the transverse correlators show that $\langle\sigma^{+}_i\sigma^{-}_{i+2}\rangle$ and $\langle\sigma^{+}_i\sigma^{-}_{i+3}\rangle$ are visibly different when comparing the complete GGE and the local GGE as $\Delta\rightarrow1$. Still, those differences are small so in experiments it might be difficult to identify which GGE is providing the correct prediction.

\subsection{Tilted ferromagnetic state}

\begin{figure}[!b]
\includegraphics[width=0.48\textwidth]{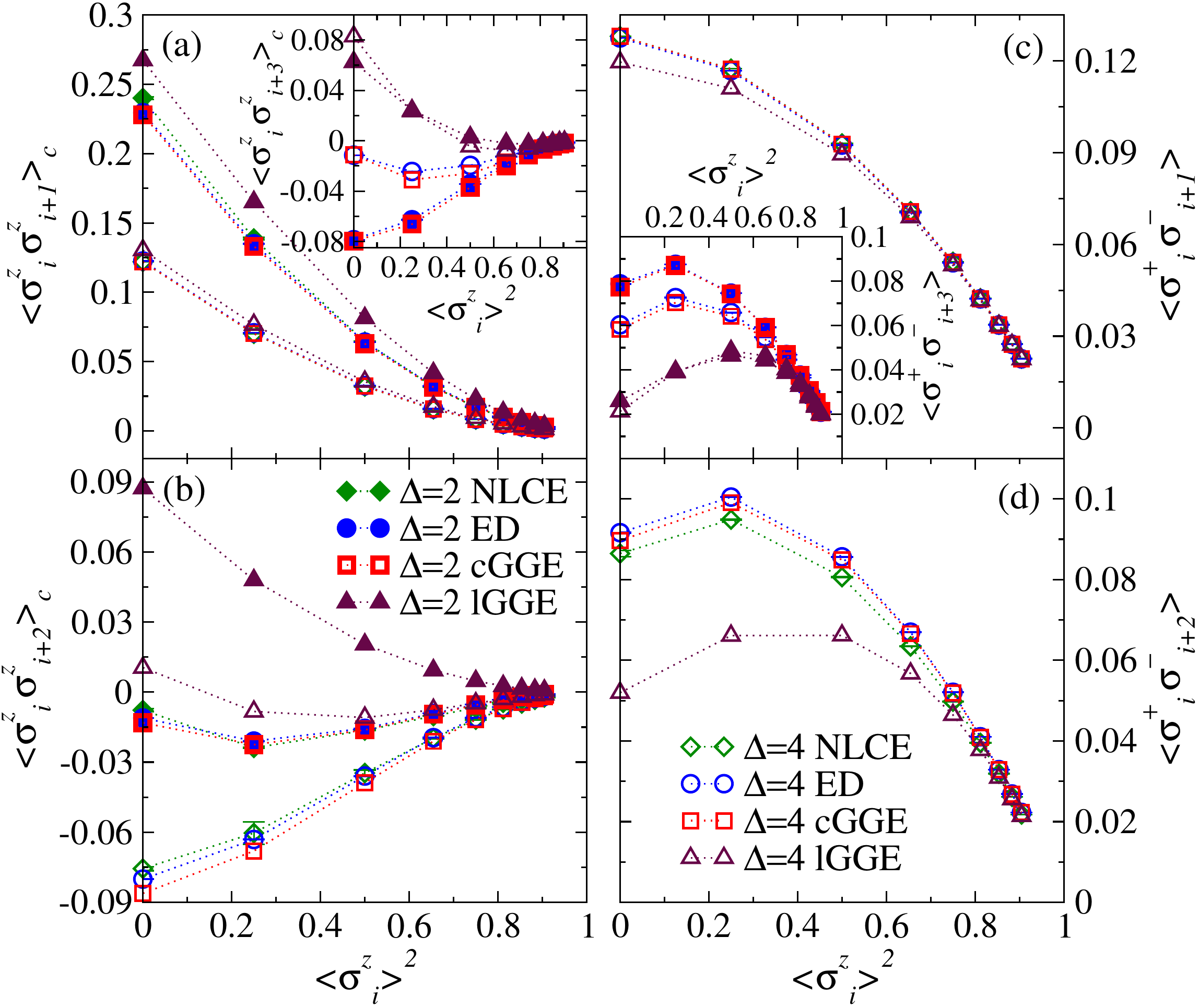}
\caption{Short-range correlators for quenches from initial tilted ferromagnetic states as functions of the squared magnetization for $\Delta=2$ and $\Delta=4$. In the main panels of (c) and (d), we only report results for $\Delta=4$. The squared magnetization is $\langle\sigma_i^z\rangle^2=\cos^2\Theta$, where $\Theta$ is the tilting angle. The complete GGE (cGGE) and local GGE (lGGE) predictions are compared to the results from ED and NLCE. The ED and NLCE results reported are the average over the two largest chains (with 23 and 24 sites) and the two highest orders of the expansion (18 and 19 orders), respectively. The values at the extremes of the errorbars depict the results that entered the average. For $\langle\sigma^{z}_i\sigma^{z}_{i+3}\rangle_{c}$, inset in (a), and $\langle\sigma^{+}_i\sigma^{-}_{i+3}\rangle$, inset in (c), we only report results for cGGE, lGGE, and ED.}
\label{fig:3}
\end{figure}

The analysis of short-range correlators in the post-quench steady state when the initial state is the tilted ferromagnet \eqref{eq:ferromagnet} reveals more interesting results. Analogously to the N\'eel state, we compute the Bethe ansatz predictions for the complete GGE and the local GGE for longitudinal and transverse correlators and compare them to ED and NLCE calculations. We consider the tilting angles:
\be
\Theta=\pi/m\,,
\label{eq:angle}
\ee
with $m=2,3,\ldots 10$. Our results are reported in Fig.~\ref{fig:3}. 

Some remarks are in order as to how the plots and calculations for the tilted ferromagnet differ from those for the N\'eel state. In the former: (i) The longitudinal correlators reported are the connected ones
\be
\langle\sigma_i^z\sigma_{i+k}^z\rangle_c=\langle\sigma_i^z\sigma_{i+k}^z\rangle-\langle\sigma_i^z\rangle^2\,,
\ee
where the squared magnetization is simply related to $\Theta$ in Eq.~\eqref{eq:angle} by the expression $\langle\sigma_i^z\rangle^2=\cos^2\Theta$. (ii) The ED results reported in the plots are obtained using the average between those obtained in chains with $L=23$ and $L=24$ sites. The results for $L=23$ and $L=24$ sites are shown as the extremes of the errorbars. 

\begin{figure}[!t]
\includegraphics[width=0.48\textwidth]{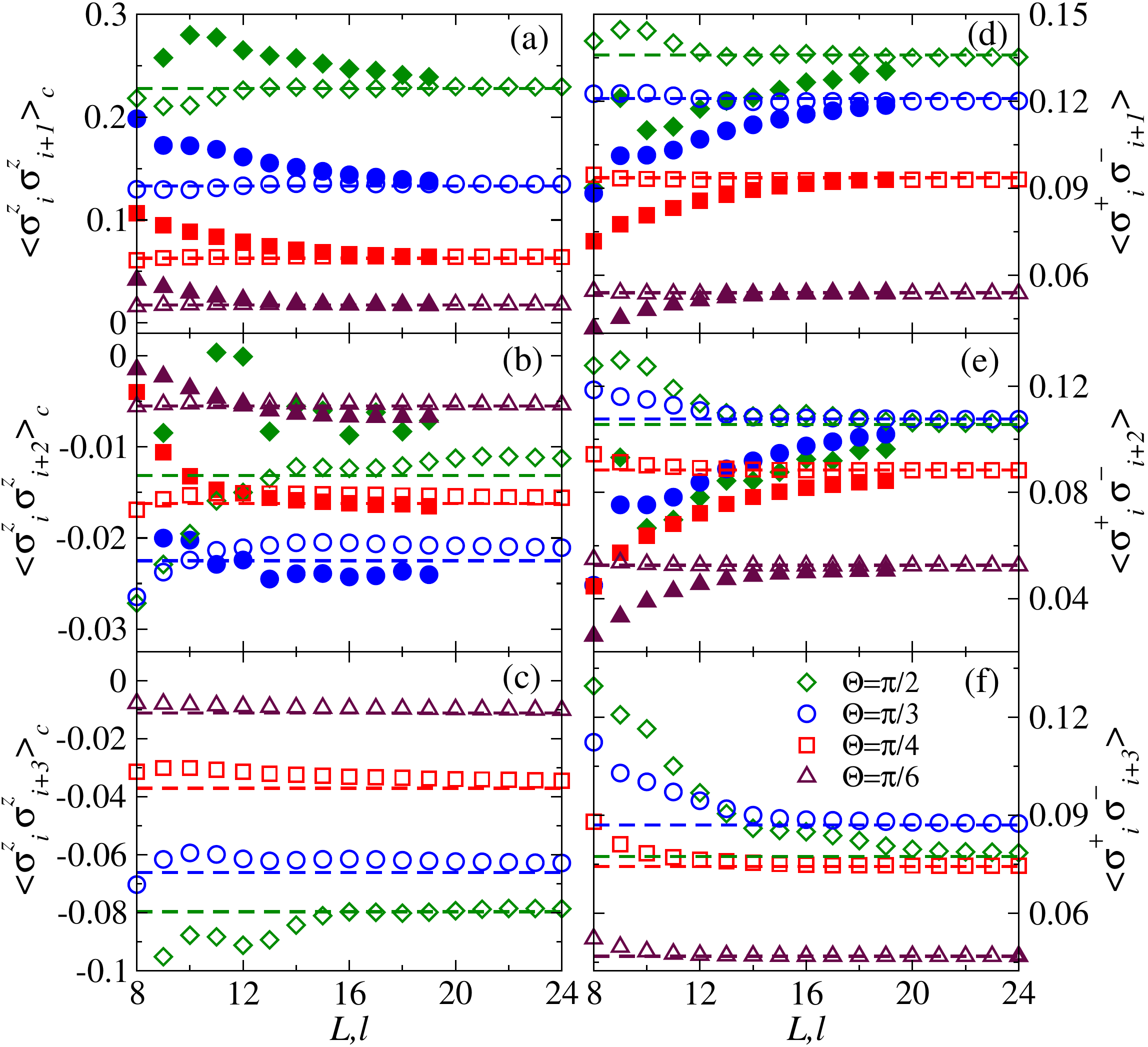}
\caption{Short-range correlators for quenches from initial tilted ferromagnetic states with tilting angles $\Theta=\pi/2,\ \pi/3,\ \pi/4,$ and $\pi/6$ for $\Delta=2$. The ED results (empty symbols) are shown as functions of the chain size $L$ in systems with periodic boundary conditions, while the NLCE results (filled symbols) are shown as functions of the order $l$ of the expansion. The complete GGE results are shown as horizontal dashed lines. For $\langle\sigma^{z}_i\sigma^{z}_{i+3}\rangle_c$ in (c), and $\langle\sigma^{+}_i\sigma^{-}_{i+3}\rangle$ in (f), we only report ED and complete GGE results. For all correlators and values of $\Theta$ shown, one can see that the ED and NLCE results approach the Bethe ansatz predictions as $L$ and $l$ increase, respectively.}
\label{fig:4}
\end{figure}

In Fig.~\ref{fig:3} one can see that, in most cases, there is an excellent agreement between the Bethe ansatz predictions and the results from ED and NLCE calculations. In the cases in which the results do not agree, we find that the exact diagonalization and NLCE calculations approach the complete GGE results with increasing the chain and the cluster sizes, respectively. In Fig.~\ref{fig:4}, we show how the ED and NLCE results converge toward the complete GGE predictions as the chain and cluster size increase, respectively. (For next-next-nearest neighbor correlators, we only show results from ED.) For this quench, we find that the ED results exhibit a faster convergence toward the complete GGE predictions than the NLCE ones. In addition, for both ED and NLCEs, the convergence worsens as the tilting angle approaches $\Theta=\pi/2$, and as the support of the correlators increases.

\begin{figure}[!t]
\includegraphics[width=0.48\textwidth]{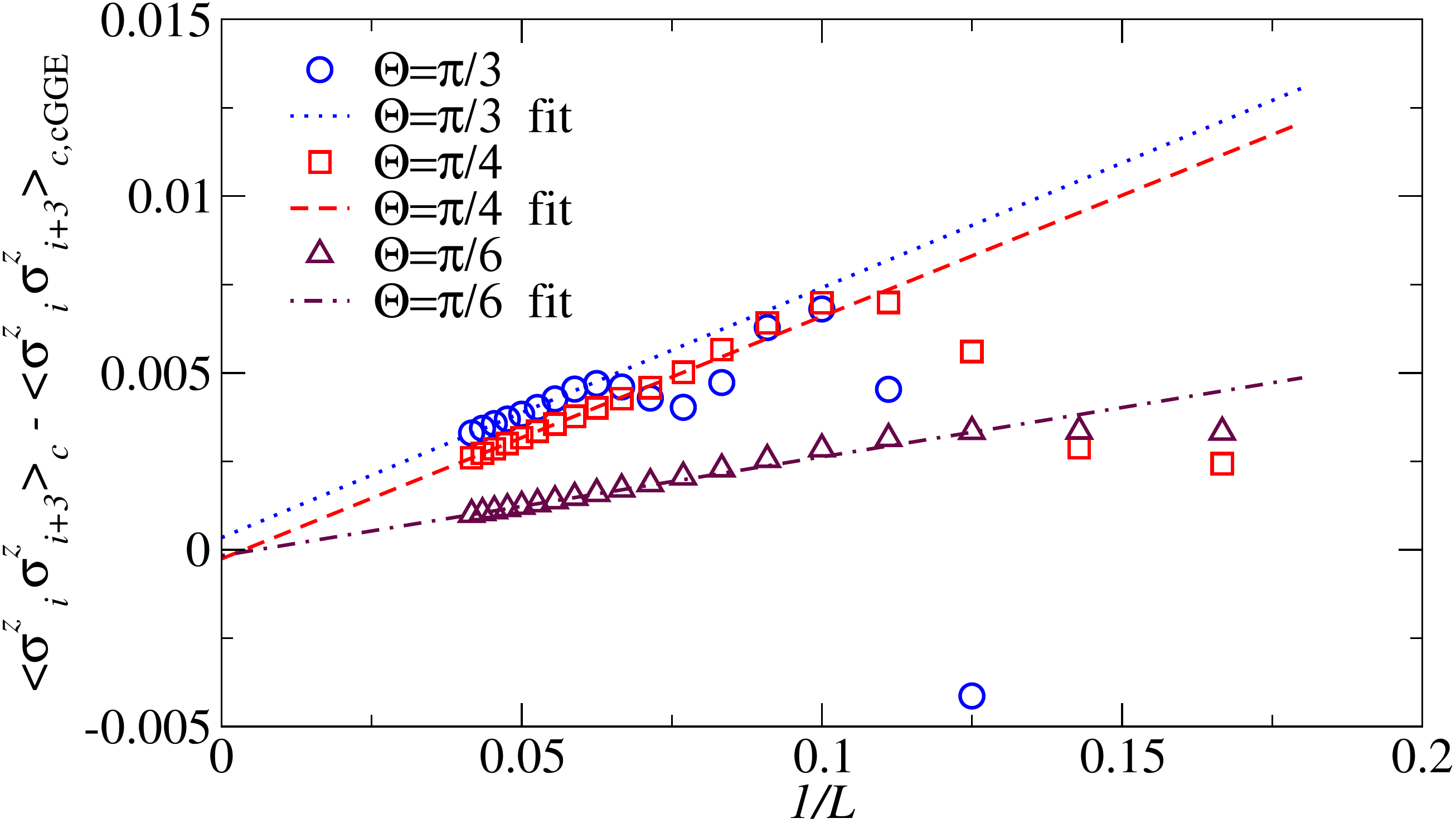}
\caption{Finite-size scaling analysis for $\langle\sigma^{z}_j\sigma^{z}_{j+3}\rangle_c$ [see also Fig.~\ref{fig:4}(c)]. Difference between ED results and the complete GGE predictions vs $1/L$ (symbols), and fits of the differences in the largest chains to $a+b/L$ (lines).}
\label{fig:4b}
\end{figure}

A detailed analysis of finite-size effects for $\langle\sigma^{z}_j\sigma^{z}_{j+3}\rangle_c$ [corresponding to the results reported in Fig.~\ref{fig:4}(c) for $\Theta=\pi/3,\ \pi/4,$ and $\pi/6$] is undertaken in Fig.~\ref{fig:4b}. We first note that the differences between the ED and complete GGE results are much smaller in Fig.~\ref{fig:4b} than in Fig.~\ref{fig:2b}. In Fig.~\ref{fig:4b}, we also report results of fits of those differences to $a+b/L$. They are consistent with a vanishing value of $a$. (We do not show results for $\Theta=\pi/2$ because finite size effects are very strong and we are not able to find a stable fitting procedure for that tilting angle.)

It is interesting to note that the results in Fig.~\ref{fig:3} show that there is a strong dependence of the (connected) short-range correlators on the tilting angle. Furthermore, we find that the dependence on the anisotropy $\Delta$ increases as the tilting angle increases (and $\langle\sigma_i^z\rangle^2$ decreases). For small tilting angles, the results for the correlators can be seen to become nearly independent of the value of $\Delta$.

The predictions of the local GGE are also shown in Fig.~\ref{fig:3}. Remarkably, they can be seen to differ significantly from those of the complete GGE. As a matter of fact, for $\langle\sigma_i^z\sigma_{i+2}^z\rangle_c$ and $\langle\sigma_i^z\sigma_{i+3}^z\rangle_c$, one can see that the local GGE even predicts the wrong sign for the correlators in the steady state. These results are in stark contrast to those starting from the N\'eel state, for which the local GGE yielded relatively accurate results. They make apparent that there is no reason for one to expect the local GGE to generically provide accurate predictions for short-range correlators after quenches in interacting integrable systems. Also, our findings for $\langle\sigma_i^z\sigma_{i+2}^z\rangle_c$ and $\langle\sigma_i^z\sigma_{i+3}^z\rangle_c$ indicate that those correlators could be used in experiments with quenches from initial tilted ferromagnetic states to confirm the correctness of the complete GGE and the failure of the local GGE. 

\section{Diagonal entropies}\label{sec:diagonal_entropies}

In the previous section, we focused on short-range correlators. While they help characterize the equilibrated state after the quench, and can be probed experimentally, there are other quantities, such as the entropy, that provide complementary information about the steady state.

The notion of entropy is a fundamental cornerstone in statistical physics. In thermal equilibrium, the von Neumann entropy
\be
S_\text{vN}[\rho]=-{\rm tr}(\rho\log\rho)\,,
\label{eq:von_neumann}
\ee
provides the correct microscopic definition for the thermodynamic entropy (using the thermal Gibbs density matrix for $\rho$). From the Bethe ansatz point of view, it is an established result that the thermal entropy computed using Eq.~\eqref{eq:von_neumann} is equal to the so-called Yang-Yang entropy
\bea
S_{\rm YY}\left[\{\rho_n\}_{n=1}^{\infty}\right]&=&\sum_{n=1}^{\infty}\int_{-\pi/2}^{\pi/2}{\rm d}\lambda\,\Big\{ \rho_n(\lambda)\log\left[1+\eta_n(\lambda)\right]\nonumber\\
&+&\rho^{h}_n(\lambda)\log\left[1+\eta^{-1}_n(\lambda)\right]\Big\}\,,
\label{eq:yang_yang}
\eea
where $\eta_n(\lambda)=\rho^{h}_n(\lambda)/\rho_n(\lambda)$, while $\rho_n(\lambda)$, $\rho^h_n(\lambda)$ are the rapidity and hole distribution functions corresponding to the thermal Gibbs ensemble, cf. Sec.~\ref{sec:bethe_ansatzs}.

When entering the realm of non-equilibrium physics, providing a good definition of entropy is less immediate \cite{dkpr-16}. In this work, we focus on the entropy of the post-quench steady state. A natural candidate is provided by the infinite-time limit of the von Neumann entropy of a finite subsystem $\mathcal{A}$, with reduced density matrix $\rho_{\mathcal{A}}(t)$, of an infinite system. This entropy is also known as the entanglement entropy and is extensive, namely, it grows linearly with the length $\ell$ of the subsystem $\mathcal{A}$ \cite{cc-05}. 

It is almost automatic to identify the von Neumann entropy of the reduced density matrix in the long-time limit with the entropy of the complete GGE. Indeed, assuming its validity, the latter gives the reduced density matrix for any finite subsystem $\mathcal{A}$ in the infinite-time limit, provided that the infinite system size limit is taken first. On the other hand, the entropy of the complete GGE is computed by means of Eq.~\eqref{eq:yang_yang}, namely, it is given by the Yang-Yang entropy of the corresponding rapidity distribution functions $\rho_n(\lambda)$ and $\rho^h_n(\lambda)$.

In Refs.~\cite{rk-06, polkovnikov-11,spr-11}, and more recently in Ref.~\cite{dkpr-16}, it was discussed that the von Neumann entropy of the diagonal ensemble [see Eq.~\eqref{eq:diagensem}], also known as the diagonal entropy,
\be
S_\text{DE}=-{\rm tr}(\rho_\text{DE}\log\rho_\text{DE})\,,
\label{eq:diagonal_entropy}
\ee 
provides the correct microscopic definition of the thermodynamic entropy for the steady state of isolated quantum systems after a quench. In particular, it was argued in Ref.~\cite{polkovnikov-11} that the diagonal entropy has the correct extensivity properties and an interpretation in terms of the logarithm of the number of microstates can be given.

The relation between the diagonal entropy and the Yang-Yang entropy associated with the GGE has been studied in several works in the literature \cite{rf-11, heri-12, spr-11, gurarie-13, ckc_2-14, fc-08, vr-16}. However, all those studies focused on systems that were either noninteracting or for which a mapping onto noninteracting ones was available. In the cases in which the systems were translationally invariant \cite{gurarie-13, ckc_2-14, fc-08, vr-16}, a simple relation between the two entropies was found, namely
\be
S_{\rm DE}=\frac{1}{2}S_{\rm YY}\,.
\label{eq:relation_entropies}
\ee
A heuristic explanation for the factor $1/2$ was provided for the transverse-field Ising chain \cite{gurarie-13}. After the quench, free fermionic excitations are created in pairs of opposite momentum. This represents a set of non-trivial correlations on the quasi-particle content of the system, which constrains the entropy. However, such correlations are absent for the (complete) GGE, as it is most easily visualized for the reduced density matrix of a finite subsystem $\mathcal{A}$. Indeed, if a quasi-particle with a given momentum is in $\mathcal{A}$, the associated quasi-particle of opposite momentum will be found outside of $\mathcal{A}$ at sufficiently long times \cite{gurarie-13}.

It is natural to question whether Eq.~\eqref{eq:relation_entropies} remains valid for fully interacting systems or if additional effects due to interactions arise. In order to answer this question for quenches to the XXZ spin-1/2 Heisenberg chain, we have computed the Yang-Yang entropy [Eq.~\eqref{eq:yang_yang}] and the diagonal entropy [Eq.~\eqref{eq:diagonal_entropy}] for the N\'eel state and the tilted ferromagnet. The Yang-Yang entropy is obtained directly using the Bethe ansatz rapidity distribution functions associated with the complete GGE, while the diagonal entropy is computed numerically using ED for chains with up to $L=24$ sites (the ED results converge faster with increasing chain sizes than the NLCE ones with increasing cluster sizes so only the former are reported). For comparison, we also computed the Yang-Yang entropy for the local GGE using Bethe ansatz and the entropy of the grand canonical ensemble using NLCEs (the NLCE results for this quantity converge faster with increasing cluster sizes than the ED ones with increasing chains sizes \cite{isr-15} so only the former are reported.) All entropies reported in this section are entropies per-site.

\begin{figure}[!t]
\includegraphics[width=0.48\textwidth]{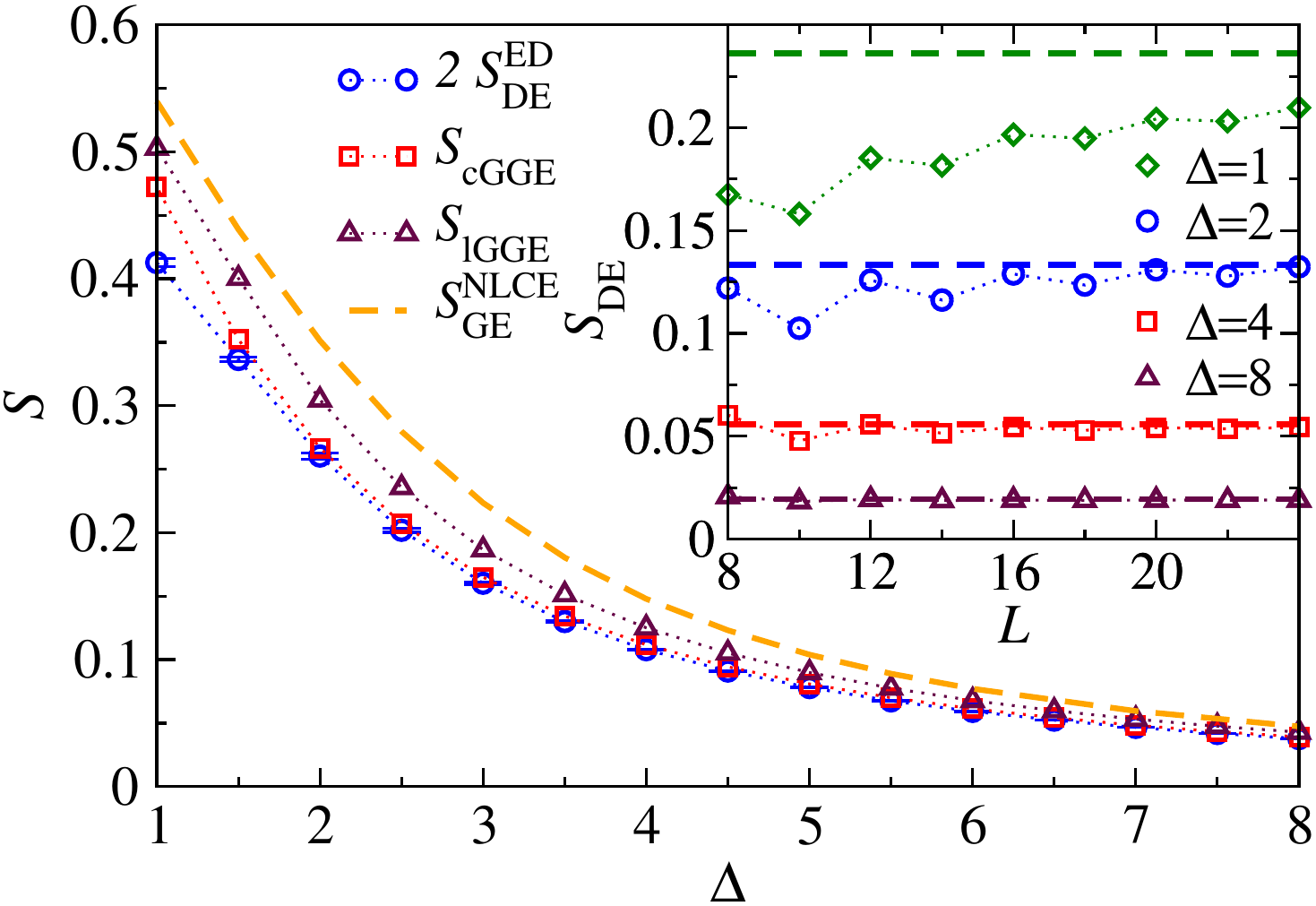}
\caption{Diagonal, grand canonical ensemble, and Yang-Yang entropies for the quench from the N\'eel state as functions of the anisotropy parameter $\Delta$. The diagonal entropy ($2S_\text{DE}^\text{ED}$, notice the factor 2) computed using ED is compared to the grand canonical ensemble entropy ($S_\text{GE}^\text{NLCE}$) computed using NLCEs, and to the Yang-Yang entropy for the complete GGE ($S_\text{cGGE}$) and for the local GGE ($S_\text{lGGE}$) computed using Bethe ansatz. The $S_\text{DE}^\text{ED}$ results reported are the average over the two largest chains (with 22 and 24 sites) considered, the extremes of the errorbars depict the results that entered the averages, while the $S_\text{GE}^\text{NLCE}$ results are for clusters with up to 18 sites and are converged to the thermodynamic limit result \cite{isr-15}. The inset shows $S_\text{DE}^\text{ED}$ and $S_\text{cGGE}/2$, notice the factor 2, for anisotropy parameters $\Delta=1,$ 2, 4, and 8. The ED results (symbols) are shown as functions of the chain size in systems with periodic boundary conditions. The complete GGE results are shown as horizontal dashed lines. For all values of $\Delta$ shown, one can see that the ED results approach the Bethe ansatz predictions as $L$ increases.
}
\label{fig:5}
\end{figure}

Our results for the N\'eel state are reported in Fig.~\ref{fig:5}. First, we note that the diagonal entropy for the N\'eel state, as computed using ED, is clearly smaller than the thermal (grand canonical ensemble) entropy, as obtained using NLCEs. This is expected as the thermal ensemble contains less information about the system than the diagonal ensemble. Analogously, the local GGE displays an entropy that is smaller than the thermal entropy but larger than the complete GGE entropy.

More importantly, we find that the entropy of the complete GGE is consistent with twice that of the diagonal ensemble. This is better seen in the inset in Fig.~\ref{fig:5}, where the ED results for the diagonal entropy are shown to approach one half of the complete GGE ones as the sizes of the chains increase. That inset also shows that finite-size effects in ED increase as the anisotropy parameter approaches the Heisenberg point $\Delta=1$. We emphasize that Eq.~\eqref{eq:relation_entropies} holds for the complete GGE and not for the local GGE, as made clear by our results in Fig.~\ref{fig:5}. The latter ensemble does not contain the information required to construct the reduced density matrix for a finite subsystem $\mathcal{A}$ of an infinite system in the infinite-time limit.

\begin{figure}[!t]
\includegraphics[width=0.48\textwidth]{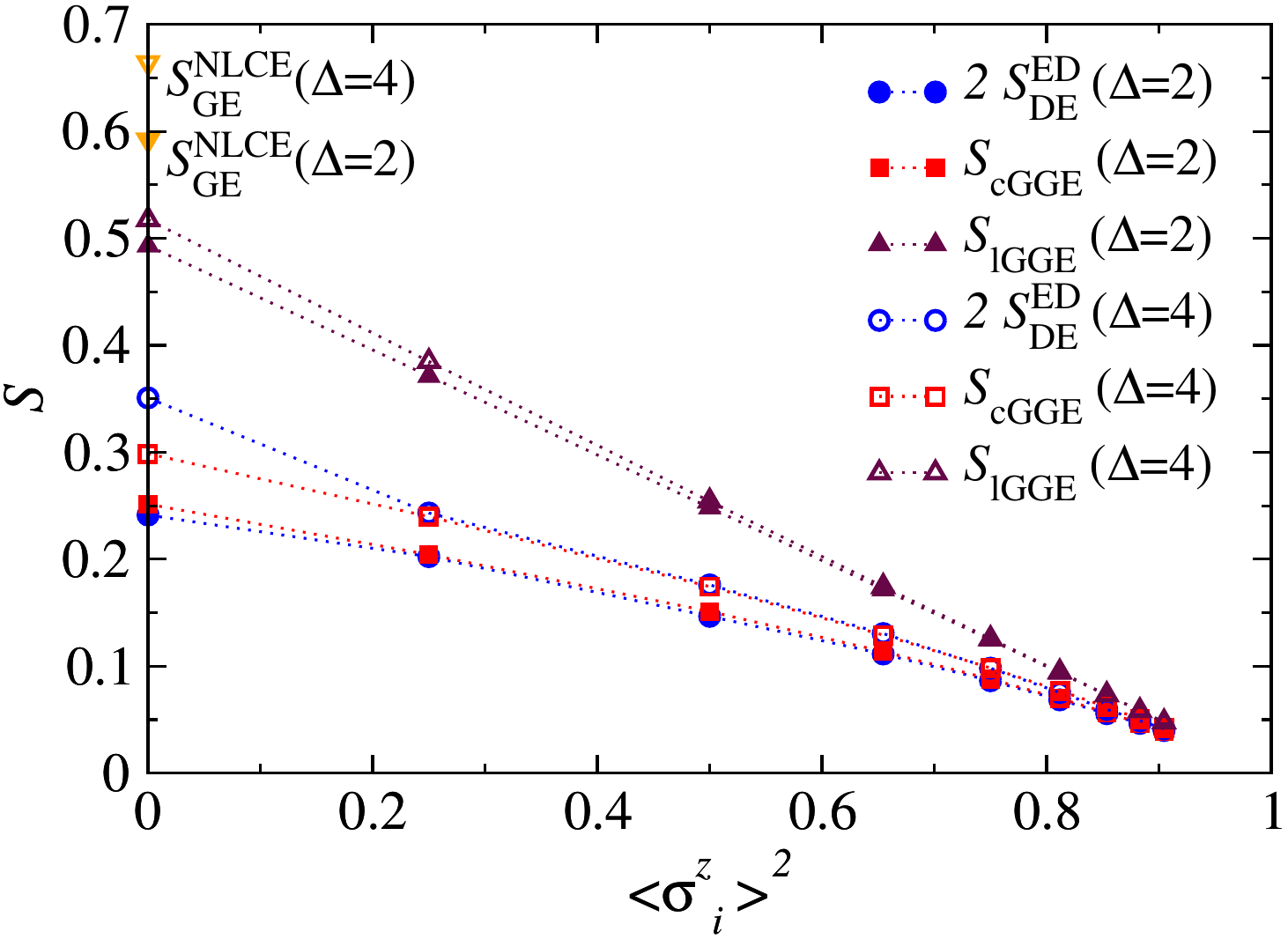}
\caption{Diagonal and Yang-Yang entropies for quenches from initial tilted ferromagnetic states as functions of the squared magnetization for $\Delta=2$ and $\Delta=4$. The diagonal entropy ($2S_\text{DE}^\text{ED}$, notice the factor 2) computed using ED is compared to the Yang-Yang entropy for the complete GGE ($S_\text{cGGE}$) and for the local GGE ($S_\text{lGGE}$) computed using Bethe ansatz. We also report NLCE results for the grand canonical ensemble entropy ($S_\text{GE}^\text{NLCE}$) for tilting angle $\Theta=\pi/2$ (zero magnetization). The results for $2S_\text{DE}^\text{ED}$ reported are obtained after an extrapolation to the thermodynamic limit, cf. appendix~\ref{app:fit}, while the $S_\text{GE}^\text{NLCE}$ results are for clusters with up to 18 sites and are converged to the thermodynamic limit result \cite{isr-15}.}
\label{fig:6}
\end{figure}

Our results for the tilted ferromagnet are displayed in Fig.~\ref{fig:6}. For this quench, we have found that finite size effects in the ED calculations are severe and, unlike for the N\'eel state, a direct comparison between the ED results for the largest chains and the complete GGE results is not meaningful. Instead, careful finite-size scaling analyses of the ED results are required. The fitting procedure followed is explained in appendix~\ref{app:fit}. Here, we focus on discussing the results obtained with it.

Figure~\ref{fig:6} shows that, like for the N\'eel state, the result for the Yang-Yang entropy in the complete GGE is consistent with twice the extrapolated ED result of the entropy in the diagonal ensemble. We find that finite-size effects in the ED calculations increase significantly as the tilting angle approaches $\Theta=\pi/2$, and the fitting procedure becomes unstable. Accordingly, we observe a discrepancy for $\Theta=\pi/2$ between the Bethe ansatz and extrapolated ED result. The discrepancy is larger for $\Delta=4$ than for $\Delta=2$. Those discrepancies are most likely a result of our fitting procedure failing to predict the diagonal entropy per site in the thermodynamic limit.

Finally, Fig.~\ref{fig:6} also shows that the Yang-Yang entropy of the local GGE is much larger than the one of the complete GGE. For $\Theta=\pi/2$ and $\Delta=2$, $S_\text{lGGE}$ is about two times larger than $S_\text{cGGE}$, and $S_\text{lGGE}$ is closer to the entropy of the grand-canonical ensemble $S^\text{NLCE}_\text{GE}$ than to $S_\text{cGGE}$. This unambiguously demonstrates that, in contrast to the N\'eel state, for the tilted ferromagnet the local GGE contains much less information than the complete GGE. For this state, neglecting the contribution of the quasi-local charges results in an ensemble that is significantly different from the one needed to describe the steady state following a quantum quench. This offers further support to the conclusions of our analysis of short-range correlators in Sec.~\ref{sec:local_correlations}. 

\section{Conclusions}\label{sec:conclusions}

We studied quantum quenches to the XXZ spin-1/2 Heisenberg chain from the N\'eel and tilted ferromagnetic states. We focused on the post-quench steady state, and presented a detailed comparison between the Bethe ansatz predictions for the complete and local GGE and numerical calculations for the diagonal ensemble by means of ED and NLCEs. 

Our analysis of short-range correlators provides one of the most accurate benchmarks to date of the validity of the complete GGE. Furthermore, we have shown that the local GGE predictions differ significantly from those of the complete GGE in the case of tilted ferromagnets. This discrepancy can be clearly and without ambiguity resolved by ED and NLCE calculations, and could potentially be tested in experiments as the equilibration times for the correlators discussed here appear to be below $20\hbar/J$ \cite{fcec-14, pvc-16}, where $J$ is the $\sigma^{x}_j\sigma^{x}_{j+1}$ and $\sigma^{y}_j\sigma^{y}_{j+1}$ coupling strength (set to 1 in this work).

Furthermore, we calculated the diagonal entropy for the complete GGE in the XXZ spin-1/2 Heisenberg chain after quantum quenches from different initial states. Using careful finite-size scaling analyzes, we found that the diagonal entropy is consistent with one half the Yang-Yang entropy for both the N\'eel and the tilted ferromagnet state. We argued that our findings are consistent with the picture of pair quasi-particle production after a quench, recovering the results obtained in several translationally invariant systems that are either noninteracting or that can be mapped onto noninteracting ones \cite{gurarie-13, ckc_2-14, fc-08, vr-16}.

\section{Acknowledgments}
We are grateful to Maurizio Fagotti and Michael Brockmann for providing us a sample of data to test our numerical code for computation of short-range correlators in the local GGE. MR is grateful to Fabian Essler for motivating discussions. This work was partially supported by the ERC under Starting Grant 279391 EDEQS (PC and EV), and by the Office of Naval Research (MR).

\appendix

\section{Exact diagonalization and \\numerical linked cluster expansions}\label{sec:diag_nlces}

The ED calculations are performed in chains with periodic boundary conditions and with up to $L=24$ sites.

To take full advantage of translational symmetry, we do not deal directly with the N\'eel state in Eq.~\eqref{eq:neel} but rather with its translational invariant version:
\be\label{eq:neelt}
|N\rangle'=\frac{1}{\sqrt{2}}(|\uparrow\downarrow\rangle^{\otimes L/2}+|\downarrow\uparrow\rangle^{\otimes L/2}),
\ee
This state can only be accommodated in chains with an even number of sites and belongs to the zero magnetization sector [the total magnetization commutes with the XXZ spin-1/2 Hamiltonian \eqref{eq:hamiltonian}]. In addition, this state belongs to the zero momentum sector and it is parity even. All our calculations for the diagonal ensemble are performed in this sub-sector of the Hilbert space. The largest matrices that need to be fully diagonalized for $L=24$ have linear dimension $D=56,822$.

The tilted ferromagnetic state in Eq.~\eqref{eq:ferromagnet} is already translationally invariant and belongs to the zero momentum -- parity even -- sector. However, in contrast to the N\'eel state, the tilted ferromagnet is a superposition of states that belong to all magnetization sectors. As a result, while the largest matrices that need to be diagonalized for each chain of size $L$ are the same size as those for the N\'eel state, one also has to diagonalize many other smaller ones for all non-zero magnetization sectors. In addition, this state can also be accommodated in chains with an odd number of sites, so we have done calculations for chains with even and odd number of sites.

The NLCE calculations for the diagonal ensemble, on the other hand, allow one to compute the expectation value of extensive observables (per lattice site, $\mathcal{O}$) after a quench in translationally invariant lattice systems in the thermodynamic limit \cite{rigol-14}. This is done by summing over the contributions from connected clusters $c$ that can be embedded on the lattice
\begin{equation}
\label{eq:LCE1}
\mathcal{O}=\sum_{c}M(c)\times \mathcal{W}_{O}(c).
\end{equation}
In the expression above, $M(c)$ is the multiplicity of cluster $c$ (the number of ways per site in it can be embedded on the lattice) and $\mathcal{W}_{O}(c)$ is the weight of the observable of interest $O$ in cluster $c$. The weight $\mathcal{W}_{O}(c)$ is computed using the inclusion-exclusion principle:
\begin{equation}
\label{eq:LCE2}
 \mathcal{W}_{O}(c)=O_\text{DE}(c)-\sum_{s \subset c} \mathcal{W}_{O}(s),
\end{equation}
where the sum runs over all connected sub-clusters of cluster $c$ and
\begin{equation}
\label{eq:LCE3}
O_\text{DE}(c)={\textrm{Tr} [O\,\rho^c_\text{DE}]}/
{\textrm{Tr} [\rho^c_\text{DE}]}
\end{equation}
is the expectation value of $O$ in the diagonal ensemble calculated for the finite cluster $c$. $\rho^c_\text{DE}$ is the many-body density matrix of the diagonal ensemble in cluster $c$. $O_\text{DE}(c)$ is computed using full exact diagonalization.

NLCEs were originally introduced to study observables for lattice systems in thermal equilibrium in the thermodynamic limit~\cite{rbs-06}. For a pedagogical introduction to NLCEs see Ref.~\cite{tkr-13}. This is the approach we used to compute the grand canonical ensemble results for the entropy ($S_\text{GE}^\text{NLCE}$) shown in Figs.~\ref{fig:5} and \ref{fig:6}. The temperature and the longitudinal magnetic field are chosen such that the energy and the magnetization per site in the grand canonical ensemble match the values set by the initial state.

In this work, NLCEs for the diagonal ensemble are carried out in clusters with up to 19 lattice sites. This means that the sum in Eq.~\eqref{eq:LCE1} only contains results for clusters with at most 19 sites. To gain an understanding of how the results converge to the thermodynamic limit ones as the cluster sizes are increased, we denote by $\mathcal{O}_l$ the sum in Eq.~\eqref{eq:LCE1} when all clusters with up $l$ sites are included. $l$ is usually referred to as the order of the expansion and is reported in the plots in Fig.~\ref{fig:4}. The fact that the clusters in NLCEs have open boundary conditions, namely, they lack translational symmetry and hence are more costly to fully diagonalize, explains why we are limited to smaller cluster sizes in NLCEs than chains sizes in ED. 

NLCEs for the N\'eel state were previously carried out in Refs.~\cite{wdbf-14, rigol-14a} for clusters with up to 18 lattice sites. Details about the NLCE calculations can be found in Ref.~\cite{rigol-14a}. Here we have extended these calculations by one order (to clusters with up to 19 lattice sites) and also computed transverse correlators. 

The NLCE calculations for the tilted ferromagnet are carried out in a similar fashion. For each cluster in the expansion, we use the fact that the tilted ferromagnet is parity even and consider all magnetization sectors (in contrast to the N\'eel state \cite{rigol-14a}). This makes the NLCE calculations for the tilted ferromagnet more costly than those for the N\'eel state. Also, for the tilted ferromagnet, the connected nearest neighbor longitudinal correlators are computed after the sub-cluster subtraction [Eq.~\eqref{eq:LCE2}] is performed for the non-connected correlators. On the other hand, for next nearest neighbors sites, the connected longitudinal correlators are computed at the cluster level and the sub-cluster subtraction is performed afterwards. This accelerates the convergence of the NLCE results. The largest matrices that need to be fully diagonalized for both initial states in clusters with $L=19$ have linear dimension $D=46,252$.

\section{Rapidity distribution functions for the local and complete GGE}\label{sec:app_distributions}
In this appendix, we review results regarding the rapidity distribution functions corresponding to the complete and local GGE both for the N\'eel and tilted ferromagnet state.

In the case of the complete GGE, the rapidity distribution functions can be determined analytically. In particular, for the N\'eel state they read \cite{wdbf-14}
\bea
\eta_1(\lambda)&=&\frac{\sin^{2}(2\lambda)\left[\cosh(\eta)+2\cosh(3\eta)-3\cos(2\lambda)\right]}{2\sin\left(\lambda-i\frac{\eta}{2}\right)\sin\left(\lambda+i\frac{\eta}{2}\right)}\nonumber\\
&\times & \left[\sin(2\lambda-2i\eta)\sin(2\lambda+2i\eta)\right]^{-1}\,,
\eea
\bea
\rho_1^{h}(\lambda)&=&a_1(\lambda)\nonumber\\
&\times & \left(1-\frac{\cosh^2(\eta)}{\pi^2a^2_1(\lambda)\sin^2(2\lambda)+\cosh^2(\eta)}\right)\,,
\eea
where $a_1(\lambda)$ is given in Eq.~\eqref{def:a_function}. Higher string distribution functions are obtained as
\be
\eta_n(\lambda)=\frac{\eta_{n-1}(\lambda+i\eta/2)\eta_{n-1}(\lambda-i\eta/2)}{\eta_{n-2}(\lambda)+1}-1\,,
\label{eq:eta_relation}
\ee
\bea
\rho^h_{n}(\lambda)&=&\rho^h_{n-1}(\lambda+i\eta/2)[1+\eta^{-1}_{n-1}(\lambda+i\eta/2)]\nonumber\\
&+&\rho^h_{n-1}(\lambda-i\eta/2)[1+\eta^{-1}_{n-1}(\lambda-i\eta/2)]\nonumber\\
&-&\rho^h_{n-2}(\lambda)\,,
\label{eq:holes_relation}
\eea
where we set $\eta_0(\lambda)\equiv 0$, $\rho_0^h(\lambda)\equiv 0$.

In the case of the tilted ferromagnet, the rapidity distribution functions corresponding to the complete GGE instead are \cite{pvc-16} 
\bea
\eta_1(\lambda)&=&-1+ 
  \frac{T_1\left( \lambda + i \frac{\eta}{2} \right)}{\phi\left( \lambda + i \frac{\eta}{2} \right)}
  \frac{T_1\left( \lambda- i \frac{\eta}{2} \right)}{\bar{\phi}\left( \lambda - i \frac{\eta}{2} \right)} \,\label{tf_eta1},
\\
 \rho^h_{1}(\lambda)&=&  \frac{\sinh\eta}{\pi } \left(\frac{1}{\cosh (\eta )-\cos (2 \lambda)}-\frac{P(\lambda)}{Q(\lambda)}\right)\,,\label{tf_rhoh1}
\eea
where 
\bea
P(\lambda)&=&2 \sin ^2(\Theta ) \Big\{2 \sin ^2(\Theta )\nonumber\\
&+&\cosh
   (\eta ) \left[(\cos (2 \Theta )+3) \cos (2 \lambda)+4\right]\Big\}\,,\\
Q(\lambda)&=&   \sinh ^2(\eta ) \left[\cos (2 \Theta )+3\right]^2 \sin ^2(2 \lambda)
+\big\{2 \sin ^2(\Theta )\nonumber\\
&+&\cosh (\eta ) \left[(\cos (2 \Theta )+3) \cos (2 \lambda)+4\right]\big\}^2\,,\\
T_1(\lambda) &=& \cos (\lambda ) \big[4 \cosh (\eta )\nonumber\\
&-&2 \cos (2 \Theta ) \sin ^2 \lambda +3 \cos (2 \lambda )+1\big]\,, \\
\phi(\lambda) &=&  2 \sin^2 \Theta \sin\lambda \cos\left(  \lambda + i \frac{\eta}{2} \right)  \sin\left(  \lambda - i \frac{\eta}{2} \right)\,, \\
\bar{\phi}(\lambda) &=&  2 \sin^2 \Theta \sin\lambda \cos\left(  \lambda - i \frac{\eta}{2} \right)  \sin\left(  \lambda + i \frac{\eta}{2} \right) \,.
\eea
Once again, higher string distribution functions are immediately given by Eqs.~\eqref{eq:eta_relation} and \eqref{eq:holes_relation}.

As opposed to the complete GGE, the rapidity distribution functions for the local GGE can only be obtained numerically. In the case of the N\'eel state, these were explicitly obtained in Ref.~\cite{wdbf-14}, where a numerical scheme was developed to this end (see also Ref.~\cite{pozsgay2-14}). This method requires an initial guess for the first rapidity distribution function $\rho_1(\lambda)$ and reaches the correct distribution functions by subsequent iterations \cite{wdbf-14}. This scheme was used here to obtain the local GGE predictions corresponding to the N\'eel and tilted ferromagnet states, as displayed in Figs.~\ref{fig:1} and \ref{fig:3} for local correlations, and in Figs.~\ref{fig:5} and \ref{fig:6} for the Yang-Yang entropies. Note that in the case of tilted ferromagnetic states one has to account for a non-vanishing magnetization, as opposed to the cases considered in Ref.~\cite{wdbf-14, pozsgay2-14}. Accordingly, one has to introduce a Lagrange multiplier $h$ in order to fix the correct magnetization. In turn, this determines the asymptotic behavior of $\eta_n(\lambda)$ for large $n$, which has to be used to truncate the infinite system of partially decoupled equations for $\eta_n$, in complete analogy with the thermal case \cite{takahashi-99}.

In all the cases considered here, we explicitly checked that the values of the short-range correlators obtained for the local GGE were in agreement with those of Ref.~\cite{fcec-14}, where a different quantum transfer matrix approach was employed.

\section{Short-range correlators from Bethe ansatz}\label{sec:app_correlators}

Here, we briefly review the formulas used in this work to compute short-range correlations by means of the Bethe ansatz. These were recently derived in Ref.~\cite{mp-14}, to which we refer the reader for further details (see also Refs.~\cite{wdbf-14,pmwk-14, pozsgay-16}).

First, we introduce the set of auxiliary functions $\{\rho^{(a)}_n(\lambda)\}_{n=1}^{\infty}$, $\{\sigma^{(a)}_n(\lambda)\}_{n=1}^{\infty}$ as the solution of the following system of integral equations
\bea
\rho_n^{(a)}(\lambda)&=&-s^{(a)}_n(\lambda)-\sum_{m=1}^{\infty}\left(\varphi_{nm}\ast\frac{\rho_m^{(a)}}{1+\eta_m}\right)(\lambda)\,,\label{rho_a_aux}\\
\sigma^{(a)}_n(\lambda)&=&\tilde{s}^{(a)}_n(\lambda)+\sum_{m=1}^{\infty}\left(\tilde{\varphi}_{nm}\ast\frac{\rho_m^{(a)}}{1+\eta_m}\right)(\lambda)\nonumber\\
&-&\sum_{m=1}^{\infty}\left(\varphi_{nm}\ast\frac{\sigma_m^{(a)}}{1+\eta_m}\right)(\lambda)\,,\label{sigma_aux}
\eea
where we use the notation in Eq.~\eqref{eq:convolution} for the convolution of two functions as well as
\bea
s^{(a)}_n(\lambda)&=&\left(\frac{\partial}{\partial\lambda}\right)^a s^{(0)}_n(\lambda)\,,\\
\tilde{s}^{(a)}_n(\lambda)&=&\left(\frac{\partial}{\partial \lambda}\right)^a \tilde{s}^{(0)}_n(\lambda)\,,\\
s^{(0)}_n(\lambda)&=&\frac{2 \sinh(n \eta)}{\cos(2 \lambda) - \cosh(n \eta)}\,,\\
\tilde{s}^{(0)}_n(\lambda)&=&-\frac{n\sin(2 \lambda)}{\cos(2 \lambda) - \cosh(n \eta)}\,,
\eea
and 
\bea
\varphi_{jk}(\lambda)&=&-\Big[(1-\delta_{jk})s^{(0)}_{|j-k|}(\lambda)+2s^{(0)}_{|j-k|+2}(\lambda)\nonumber\\
&+&\ldots+2s^{(0)}_{j+k-2}(\lambda)+s^{(0)}_{j+k}(\lambda)\Big]\,,
\\
\tilde{\varphi}_{jk}(\lambda)&=&-\Big[(1-\delta_{jk})\tilde{s}^{(0)}_{|j-k|}(\lambda)+2\tilde{s}^{(0)}_{|j-k|+2}(\lambda)\nonumber \\
&+&\ldots+2\tilde{s}^{(0)}_{j+k-2}(\lambda)+\tilde{s}^{(0)}_{j+k}(\lambda)\Big]\,. 
\eea

The short-range correlators are then given in terms of algebraic expressions  of the form 
\bea
\langle\sigma_{1}^z\sigma_{2}^z\rangle&=&\coth(\eta)\omega_{00}+W_{10}\,,\label{eq:algebraic1}\\
\langle\sigma_{1}^x\sigma_{2}^x\rangle&=&-\frac{\omega_{00}}{2\sinh(\eta)}-\frac{\cosh(\eta)}{2}W_{10}\,,\label{eq:algebraic2}
\eea
where the parameters $\omega_{ab}$ and $W_{ab}$ are defined by
\bea
\omega_{ab}&=&-(-1)^{(a+b)/2}\Omega_{ab}\nonumber\\
&-&(-1)^b\frac{1}{2}\left(\frac{\partial}{\partial \lambda}\right)^{a+b}K(\lambda)\Big|_{\lambda=0}\,,
\eea
\bea
W_{ab}&=&-(-1)^{(a+b-1)/2}\Gamma_{ab}\nonumber\\
&+&(-1)^b\frac{1}{2}\left(\frac{\partial}{\partial \lambda}\right)^{a+b}\tilde{K}(\lambda)\Big|_{\lambda=0}\,.
\eea
Here
\bea
\Omega_{ab}&=&-2\sum_{n=1}^{\infty}s^{(b)}_n\cdot\frac{\rho^{(a)}_n}{1+\eta_{n}}\,,
\eea
\bea
\hspace{-0.5cm}\Gamma_{ab}&=&2\left(\sum_{n=1}^{\infty}\tilde{s}_n^{(b)}\cdot \frac{\rho^{(a)}_n}{1+\eta_n}+\sum_{n=1}^{\infty}s_n^{(b)}\cdot \frac{\sigma^{(a)}_n}{1+\eta_n}\right)\,,
\eea
and we introduced the notation
\be
f\cdot g=\int_{-\pi/2}^{\pi/2}{\rm d}\mu f(\mu)g(\mu)\,,
\ee
and
\bea
K(\lambda)&=&\frac{\sinh(2\eta)}{\sinh(\lambda+\eta)\sinh(\lambda-\eta)}\,,\\
\tilde{K}(\lambda)&=&\frac{\sinh(2\lambda)}{\sinh(\lambda+\eta)\sinh(\lambda-\eta)}\,.
\eea
Increasing the range of the correlators, the algebraic expressions analogous to Eqs.~\eqref{eq:algebraic1} and \eqref{eq:algebraic2} become increasingly long and are not reported here. See Ref.~\cite{mp-14} for details.

Note that for transverse correlators, we compute $\langle\sigma_j^{+}\sigma_{j+k}^{-}\rangle$ as displayed in Figs.~\ref{fig:1} and \ref{fig:3}. These can be easily related to the correlators $\langle\sigma_j^{x}\sigma_{j+k}^{x}\rangle$. Indeed, we have
\bea
\sigma_j^{+}\sigma_{j+k}^{-}&=&\frac{1}{4}\big[\sigma_j^x\sigma_{j+k}^x+\sigma_j^y\sigma_{j+k}^y\nonumber\\
&-&i(\sigma^x_j\sigma^y_{j+k}-\sigma^y_j\sigma^x_{j+k})\big]\,.
\eea
Next, we note that 
\be
\langle \sigma^x_j\sigma^y_{j+k}-\sigma^y_j\sigma^x_{j+k}\rangle=0\,,
\ee
when computed on an ensemble invariant under spin-inversion, as are the local and complete GGE. Exploiting rotational invariance along the $z$-axis, we then have
\be
\langle \sigma_j^{+}\sigma_{j+k}^{-} \rangle=\frac{1}{2} \langle \sigma_j^{x}\sigma_{j+k}^{x} \rangle\,.
\ee

Equations \eqref{rho_a_aux} and \eqref{sigma_aux} can also be cast in partially decoupled form \cite{takahashi-99, mp-14}. However, for the tilted ferromagnetic state for small values of the tilting angle $\Theta$, the rapidity distribution functions of the complete GGE are peaked around $\pm\pi/2$ \cite{pvc-16}. Accordingly, we found it more convenient to solve the coupled form Eqs.~\eqref{rho_a_aux} and \eqref{sigma_aux} using the Gaussian quadrature method, which reduced the numerical error due to discretization. Conversely, for the local GGE for the tilted ferromagnet, and for the local and complete GGE for the N\'eel state, the decoupled form of these equations was used.

\section{Finite-size analysis of the diagonal entropies}\label{app:fit}

\begin{figure}[!b]
\includegraphics[width=0.48\textwidth]{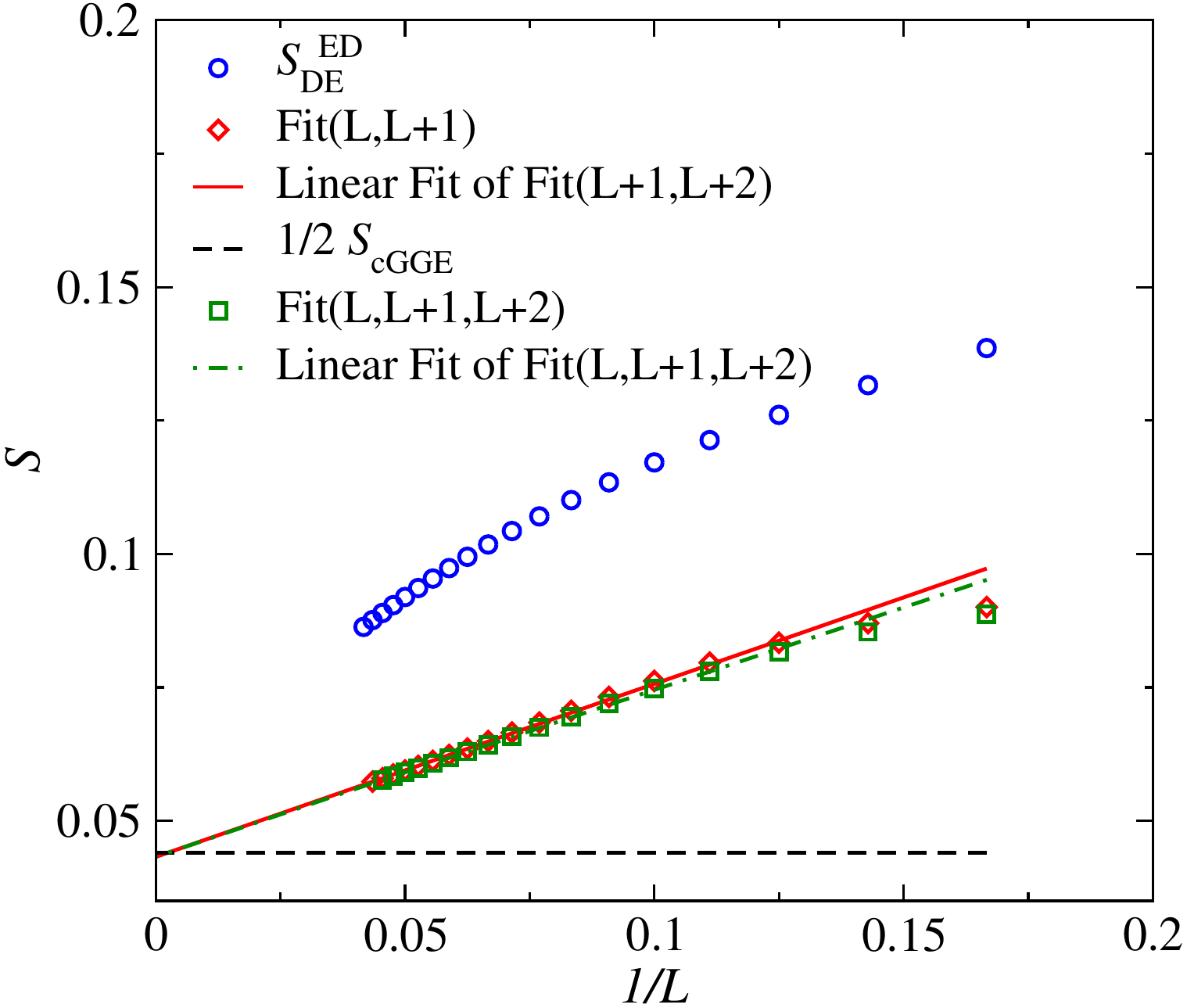}
\caption{Diagonal entropy as obtained using exact diagonalization and the extrapolations explained in the text for $\Delta=2$ and $\Theta=\pi/6$. One half of the Yang-Yang entropy for the corresponding complete GGE ($1/2\, S_\text{cGGE}$) is depicted as a horizontal line. Notice that the two extrapolations of the improved sequences are almost identical to $1/2\, S_\text{cGGE}$.}
\label{fig:7}
\end{figure}

In this appendix, we briefly discuss the strategy followed to extrapolate the ED results for the diagonal entropy of the tilted ferromagnetic state to the thermodynamic limit, which is illustrated in Fig.~\ref{fig:7}. It is based on the idea of replacing the sequence of finite-size results $S_\text{DE}(L)$ with auxiliary sequences that exhibit faster convergence. This is analogous, in spirit, to the approach developed in Ref.~\cite{vbs-79} (see also \cite{hb-81}).

From the sequence of finite-size results $S_\text{DE}(L)$ for the diagonal entropy ($S_\text{DE}^\text{ED}$ in Fig.~\ref{fig:7}), we construct a second sequence Fit$(L,L+1)$ obtained by interpolating the results for $S_\text{DE}(L)$ and $S_\text{DE}(L+1)$ with a linear fit in $L^{-1}$ and extrapolating to $L \to \infty$. Namely, 
\be
{\rm Fit}(L,L+1) 
= (L+1)S_\text{DE}(L+1)-  L S_\text{DE}(L) \,. 
\ee
The choice of a linear fit in $L^{-1}$ is justified a posteriori by the fact that the sequence of numbers Fit$(L,L+1)$ shows an almost linear dependence in $1/L$ for $L$ not too small. The new sequence, ${\rm Fit}(L,L+1)$, is then extrapolated to $L\to \infty$ using a linear fit for the largest values of $L$ ($L \geq 8$ in Fig.~\ref{fig:7}). The result of that extrapolation is what we report for the diagonal entropy in Fig.~\ref{fig:6}.

We have found this fitting procedure to be stable under small modifications. For instance, we verified that changing the intermediate sequence of two-point fits to a sequence built out of three point fits does not significantly change the final result (see Fig.~\ref{fig:7}). In contrast, direct polynomial fits of $S_\text{DE}(L)$ in powers of $L^{-1}$ were found to be highly sensible to the degree of the polynomial, as well as to the number of data points, used. However, for tilting angle $\Theta=\pi/2$, finite-size effects appear to be too strong and our fitting procedure does not provide a reliable extrapolation to the thermodynamic limit. Larger system sizes need to be calculated to obtain a good estimate of the diagonal entropy for $\Theta=\pi/2$ in the thermodynamic limit.

\end{document}